\documentclass[12pt]{article}
\usepackage{amsmath,amssymb,amsfonts,bm,marginnote,cite,graphicx,slashed,color,subcaption}
\usepackage[colorlinks=true, linkcolor=blue, citecolor=blue, linktoc=all]{hyperref}
\usepackage[usenames,dvipsnames]{xcolor}

\newcommand{\thistitle}{
On the Non-commutativity of Closed String Zero Modes}
\newcommand{\addresspi}{
	Perimeter Institute for Theoretical Physics, 
	31 Caroline St. N.,  Waterloo ON, N2L 2Y5, Canada
	}
\newcommand{\addressuiuc}{
	Department of Physics, University of Illinois,
 	1110 West Green St., Urbana IL 61801, U.S.A.
	}
\newcommand{\addressvt}{
	Department of Physics, Virginia Tech,  
	Blacksburg VA 24061, U.S.A.
	}
\newcommand{\emaillf}{lfreidel@perimeterinstitute.ca}
\newcommand{\emailrgl}{rgleigh@illinois.edu}
\newcommand{\emaildm}{dminic@vt.edu}

\newcommand{\be}{\begin{equation}}
\newcommand{\ee}{\end{equation}}
\newcommand{\beq}{\begin{eqnarray}}
\newcommand{\eeq}{\end{eqnarray}}
\newcommand{\bea}{\begin{eqnarray}}
\newcommand{\eea}{\end{eqnarray}}
\newcommand{\beqn}{\begin{eqnarray}}
\newcommand{\eeqn}{\end{eqnarray}}

\newcommand{\X}{\mathbb{X}}

\newcommand{\K}{\mathbb{K}}

\newcommand{\Pm}{\mathbb{P}}

\def\pa{\partial}

\newcommand{\rd}{\mathrm{d}}

\def\s{\sigma}

\def\s{\sigma}

\def\tx{\tilde{x}}

\def\tk{\tilde{k}}

\def\tp{\tilde{p}}
\newcommand{\tR}{{\tilde R}}
\newcommand{\myfig}[3]{
	\begin{figure}[h]
	\centering
	\includegraphics[width=#2cm]{#1}\caption{{\small #3}}\label{fig:#1}
	\end{figure}
	}

\begin{document}

\title{\thistitle}
\author{
	{Laurent Freidel$^{a}\footnote{\emaillf}$, Robert G. Leigh$^{b}\footnote{\emailrgl}$\ and Djordje Minic$^{c}$\footnote{\emaildm}}\\
	\\
	{\small ${}^a$\emph{\addresspi}}\\ 
	{\small ${}^b$\emph{\addressuiuc}}\\ 
	{\small ${}^c$\emph{\addressvt}}\\
\\}
\maketitle\thispagestyle{empty}
\vspace{-5ex}
\begin{abstract}
We explore several consequences of the recently discovered intrinsic non-commutativity of the zero-mode sector of closed string theory. In particular, we illuminate the relation between T-duality and this intrinsic non-commutativity and also note that there is a simple closed string product, equivalent to the splitting-joining interaction of the pants diagram, that respects this non-commutativity and is covariant with respect to T-duality. We emphasize the central role played by the symplectic form $\omega$ on the space of zero modes.
Furthermore, we begin an exploration of new non-commutative string backgrounds.
In particular, we show that a constant non-geometric background field leads to a non-commutative space-time. We also comment on the non-associativity that consequently arises in the presence of non-trivial flux. In this formulation, the $H$-flux as well as the `non-geometric' $Q$-, $R$- and $F$-fluxes are 
simply the various components of the flux of an almost symplectic form.
\end{abstract}
\bigskip

\setcounter{footnote}{0}
\renewcommand{\thefootnote}{\arabic{footnote}}

\section{Introduction}

One of the hallmarks of string theory, as currently understood, is its compatibility with an effective field theory description at low energy, which can be found in any standard textbook exposition \cite{Polchinski:1998rq}. In addition, there are many backgrounds for string theory giving rise to effective field theories in a variety of space-time dimensions, with a wide variety of gauge interactions. Implicit in all of these constructions is the assumption that string theory behaves itself, reducing to ordinary local field theories. On the other hand, we know that this is at best a truncation, because of the many remarkable properties of string theory, such as its dualities. There is nothing sacrosanct about one particular construction, including the nature of space-time itself.

Recently \cite{Freidel:2017wst} we have uncovered an intrinsic non-commutativity in closed string theory. 
In this paper, we explore some of the implications of this result.
The non-commutativity appears in the simplest compactification of all: toroidal compactifications with no background fields. 
We related the non-commutativity to four different phenomena: we showed it was due to the presence of 
certain co-cycle factors in the operator algebra; we identified it as the requirement of causality of the commutator algebra of the string coordinates; we related it to the presence of edge modes that appear on the seams of the compactified string when it is unwrapped in its universal cover; and we related it to the presence of a coupling of the string to topological modes encoded in a  symplectic flux. 

Here we explore additional consequences of the non-commutativity.
First, we trace the presence of non-commutativity to familiar but non-trivial properties of T-duality. From this point of view, these properties follow directly within a simple (non-commutative)  operator representation, whereas previously they were understood only as a result of tracking certain operatorial `fudge factors'.

To be precise, there are two notions of co-cycle used in this context that we should be careful to disambiguate. The first is algebraic and physical, required by causality and locality of the worldsheet field theory. This is the co-cycle that we called $\epsilon(\K,\K')$ in \cite{Freidel:2017wst}, and represents the phase appearing in the definition  of a Heisenberg group. The second notion of a cocycle is associated with a representation of the algebra; this notion of co-cycle has been eliminated by the recognition of the non-commutativity of the zero modes. 
 Indeed, as we showed in \cite{Freidel:2017wst}, 
one of the main benefits of the non-commutative interpretation is in the elimination of these fudge factors, resulting in a simple (non-commutative) geometric interpretation.

In toroidal backgrounds for closed string theory, it is known that T-duality is not only a property of the spectrum of string excitations, but of interactions as well. In this paper, we also construct a closed string product, representing the basic string (pants) interaction cut along its seams. We show explicitly that its form is uniquely determined by the non-commutative phase and that it is manifestly consistent with T-duality. In a sense, this can be understood as a closed string analog of the construction found in the open string sector \cite{Douglas:1997fm}. What we find is that this closed string product carries a `$\pi$-flux' of the symplectic form $\omega$; the pants interaction diagram with arbitrary states on each leg simply forms a representation of the associated Heisenberg group, fully consistent with the vertex operator algebra.

Finally, we discuss  how the introduction of non-trivial background fields is organically included into a deformation of the intrinsic non-commutative structure of the closed string. In particular, constant background fields can be turned on by simply performing $O(d,d)$ transformations, which act linearly on the symplectic form $\omega$, and thus modify the commutation relations of the zero mode fields. As an example, we consider a constant non-geometric ($\beta$) background field and show that it leads to a non-commutative space-time, and we also comment on the non-associativity that arises in the presence of non-trivial fluxes. 

\section{Non-commutativity}

To begin, we briefly review the result of \cite{Freidel:2017wst}, 
which applies to perturbative closed string theory compactified on a torus. Classically, the most general solutions  are parametrized as
\beq\label{Xsol}
X(\tau,\sigma) = X_L(\tau-\sigma) + X_R(\tau+\sigma) ,
\eeq
with
\beqn
X_L(\tau-\sigma) &=& x_L + \frac{\alpha'}{2} p_L (\tau-\sigma) +i \lambda \sum_{m= - \infty}^{\infty}  \frac{1}{m}\alpha_m e^{-im(\tau-\sigma)}
\\
X_R(\tau+\sigma) &=& x_R + \frac{\alpha'}{2} p_R (\tau+\sigma) +i \lambda \sum_{m= - \infty}^{\infty} \frac{1}{m}\tilde{\alpha}_me^{-im(\tau+\sigma)},
\eeqn
where the string length scale is denoted $\lambda\equiv \sqrt{\frac{\hbar\alpha'}{2}}$. In the compactified theory, it is natural to introduce also the dual field
\beq\label{tXsol}
\tilde{X}(\tau,\sigma)=X_R(\tau+\sigma)-X_L(\tau-\sigma).
\eeq
A careful analysis of the symplectic structure reveals that the oscillators satisfy the usual commutation relation
$
[\hat\alpha_n^a,\hat\alpha_m^b ]=   n h^{ab} \delta_{n+m}$ where $\hat\alpha^\dagger_n = \hat\alpha_{-n}$, 
and similarly for $\tilde{\alpha}$, with $h$ denoting the space-time Lorentz metric, while surprises appear in the zero-mode sector. First, the modes $x_L,p_L,x_R,p_R$ become independently dynamical due to an edge effect: the algebra between the space and momentum variables are found to be 
\beq
\left[x^a,p_b\right]=i\hbar \delta^a{}_b,\qquad \left[\tx_a,\tp^b\right]=i\hbar \delta_a{}^b .\label{xpcomm}
\eeq
In addition, there is an unexpected contribution to the symplectic form, equivalent to the  commutator
\beqn
\left[x^a,\tx_b\right]=2\pi i\lambda^2\delta^a{}_b,
\label{xtxcomm}
\eeqn
and all the other commutators vanish.
Here we have defined, following the standard notation, $x^a=x_R^a+x_L^a$, $\tx_a=h_{ab}(x_R^b-x_L^b)$ and $p_a=\frac12 h_{ab}(p_R^b+p_L^b)$, $\tp^a=\frac12(p_R^a-p_L^a)$. 

Given this notation we have shown in \cite{Freidel:2017wst} that these commutators can be rewritten as canonical worldsheet field commutation relations 
\be
\left[\hat{X}(\tau,\sigma_1),\hat{\tilde{X}}(\tau,\sigma_2)\right] =2i\lambda^2\Big[ \pi-\theta(\sigma_{12})\Big] ,
\ee
where $\sigma\in [0,2\pi]$, and $\theta(\sigma)$ is the staircase distribution with $\theta(\sigma)=\pi$ for all $\sigma$ in the open interval $\sigma\in (0,2\pi)$. 
This non-commutativity can be interpreted as an integration constant\footnote{See also \cite{Blair:2013noa} where this possibility was first discussed but not fully acted upon.} obtained by integrating the canonical equal-time commutator 
 \be
 [\hat{X}(\tau,\sigma_1),\pa_\tau\hat{{X}}(\tau,\sigma_2)]=[\hat{X}(\tau,\sigma_1),\pa_\s\hat{\tilde{X}}(\tau,\sigma_2)] = 2\pi i\hbar \alpha' \delta(\s_{12}),
 \ee with respect to $\sigma_2$. It turns out that the only value of this integration constant consistent with worldsheet causality is $\pi\alpha'\hbar$, which leads to (\ref{xtxcomm}). In addition, given the commutator (\ref{xtxcomm}), the vertex operator algebra satisfies mutual locality without the need for operatorial co-cycle factors. Indeed a representation is given simply by Weyl operators formed from $\hat{x}^a$ and $\hat\tx_a$. Although this may seem like a technicality, we will show elsewhere that it allows for a deep understanding of `non-geometric backgrounds', such as asymmetric orbifolds \cite{Narain:1986qm} and T-folds \cite{Hull:2006va}. 

A compact way \cite{Freidel:2017xsi, Freidel:2015pka} to package these commutators together is to introduce double field notation  $\X^A(\tau,\sigma)=(X^a(\tau,\sigma), \tilde{X}_a(\tau,\sigma))$, for which the above canonical commutators appear as
\beq
\left[ \hat\X^A(\tau,\sigma_1),\hat\X^B(\tau,\sigma_2)\right]
=2i\lambda^2
\left[ \pi \omega^{AB}-\eta^{AB}\theta(\sigma_{12})\right] ,
\eeq
where $\omega$ is an invertible two form and $\eta$ is a symmetric form with signature $(d,d)$. 
In \cite{Freidel:2017wst}, 
we showed that $\omega_{AB}$ should be thought of as an intrinsic part of the formulation of the Polyakov path integral. In particular, including it allows for covariance with respect to $O(d,d)$, in which $\eta_{AB}$ is invariant but $\omega_{AB}$ is not. Importantly, as stated above, in the zero mode sector, vertex operators can be thought of as involving Weyl operators which are exponentials of $\hat{\X}^A=(\hat{x}^a,\hat{\tx}_a)$ alone (that is, independent of the conjugate operators $\hat\Pm_A$), and $\omega_{AB}$ can be thought of as a symplectic form on the reduced space coordinatized by $\X^A$, a subspace of the full phase space. Thus the non-commutativity of the zero modes takes a simple form, being simply a Heisenberg algebra satisfied by $(x^a, \tx_b)$, with the string length setting the scale for the commutator. The presence of $\omega_{AB}$ in general can be summarized as the inclusion of a factor $e^{i\int\omega}$  in the Polyakov path integral. As we have emphasized elsewhere \cite{Freidel:2015pka}, the Polyakov path integral can be written in double space notation (which we refer to as metastring theory), and in this formulation, $\eta_{AB}$ and $\omega_{AB}$ play fundamental geometric roles, along with a third symmetric form $H_{AB}$. In fact, these three structures describe a (flat) {\it Born geometry}. \cite{Freidel:2013zga}\footnote{The role of the symplectic structure in the context of T-duality has also been
emphasized in \cite{Alvarez:2000bh, Alvarez:2000bi}.}

We note in passing here that although we are using the same notation often used in 
{\it double field theory} \cite{Tseytlin:1990nb, Tseytlin:1990va, Siegel:1993th, Siegel:1993xq, Siegel:1993bj, Hull:2009mi}, 
we are making significant departures in accounting for the intrinsic non-commutativity by including $\omega_{AB}$. 
Its inclusion in double field theory results in significant simplifications. 
Indeed, we will comment on some interesting aspects of this structure in the final section of this paper.
We expect that similar structure will be present in supersymmetric versions as well. 

One of the points we would like to elaborate in the following section is the role that this non-commutativity plays in T-duality. So let us begin by reviewing that notion. For a toroidal background, there are worldsheet constraints on the spectrum of the theory, which take the form
\beq
\frac{m^2}{\hbar^2}  = \left( \frac{n}{R}\right)^2 + \left(\frac{w}{\tR}\right)^2 + \frac{ N_L+{N}_R-2}{\lambda^2},
\qquad \frac{n}{R}\frac{w}{\tR} =\frac{N_L-{N}_R}{2\lambda^2},
\eeq
where $m$ is the invariant mass in non-compact Minkowski space-time and we have for simplicity of notation taken a single compact dimension (this can be generalized to higher dimensional tori without difficulty). Here $n/R$ (with $n\in\mathbb{Z}$) is an eigenvalue of $\hat{k}$, while $w/\tR$ (with $w\in\mathbb{Z}$) is an eigenvalue of $\hat{\tk}$. We have introduced what we will refer to as the {\it dual radius}, $\tR$, which satisfies \[R\tR=2\lambda^2.\] Thus the radius and dual radius are inversely proportional.\footnote{We notice that (\ref{Xsol}-\ref{tXsol}) imply that taking $X(\tau,\sigma)$ to $X(\tau,\sigma+2\pi)$ corresponds to $(x,\tx)\mapsto (x+2\pi wR,\tx+2\pi n\tR)$. Thus a state labelled by $(n,w)$ corresponds to a string wound $w$ times around $x$ and $n$ times around $\tx$.}  T-duality is the statement that this spectrum of states (as well as all other aspects of the theory) is invariant under the exchange of $(n,R)$ with $(w,\tR)$. In terms of the string zero modes, T-duality can be regarded as the map $(x_L,x_R)\mapsto (-x_L,x_R)$, or equivalently, $x\leftrightarrow \tx$. It is well known that $R\to\infty$ corresponds to decompactification where the $\tilde{x}$ mode decouples and the effective description can be achieved in terms of space-time fields $\Phi(x)$.
Then as it is often said, the limit $R\to 0$ also results effectively in decompactification, in which $x$ decouples and an effective description can be achieved in terms of dual-space-time fields $\tilde\Phi(\tilde{x})$. 
Consequently, which compact coordinate, $x$ or $\tx$, plays the role of a spatial coordinate depends on context. It is crucial to note that in each limit, a notion of locality for the  effective field interactions is recovered and what will be of interest to us is to explore the mechanism behind the appearance of the dual locality.
In fact, what we will show at the level of quantum states, is that T-duality can be regarded precisely as a certain transform between distinct bases. It is the non-commutativity of $x$ and $\tx$ that offers this interpretation. We will also show that even in the absence of the decompactification limit there is a principle that generalizes  locality in its organization of string interactions.

\section{T-duality and the role of non-commutativity}\label{sec:Tdual}

To begin, we focus on the zero-mode sector, and consider fields which we write as $\Phi_w(x)$, where $w$ denotes winding as above. It can be interpreted as a wavefunctional in the worldsheet theory, as described in Fig. \ref{fig:disk.pdf}.
\myfig{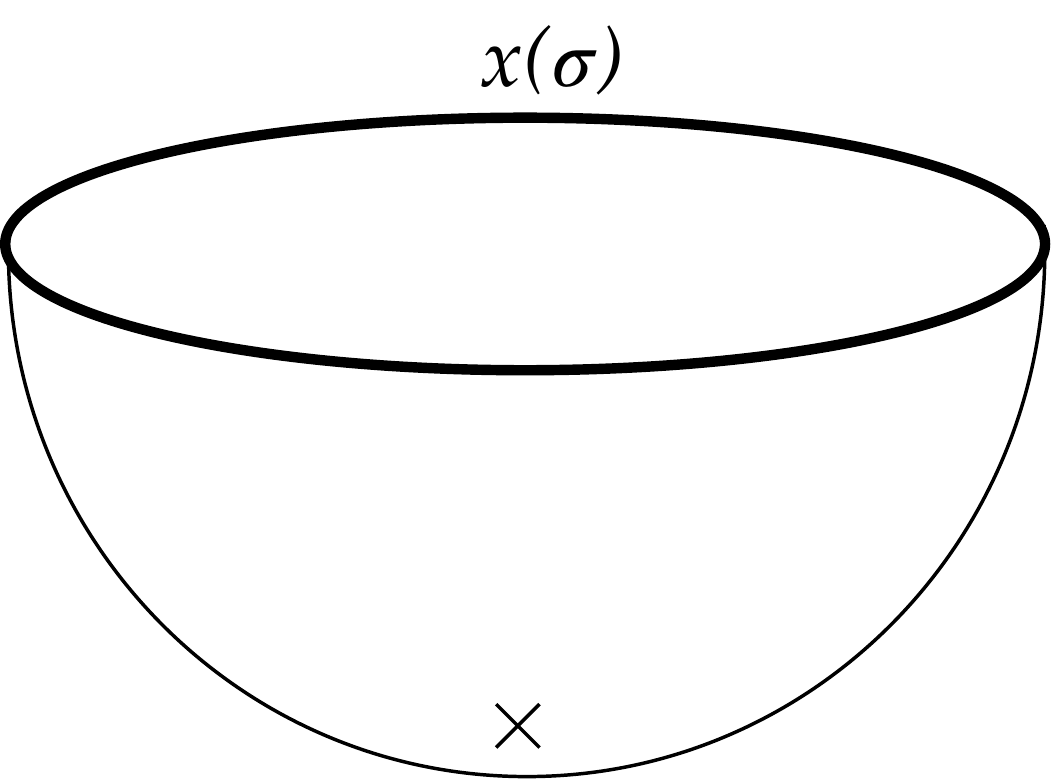}{3}{We interpret $\Phi_w(x)$ as a worldsheet wavefunctional, which can be visualized as a disk diagram with insertion corresponding to $\Phi$, and a fixed boundary  embedding labeled by $x(\sigma)$. In the current notation, this embedding is described by the zero mode $x$ and winding number $w$.}
We will find it convenient to interpret $x$ as a coordinate in the covering space. $\Phi_w(x)$ is periodic, satisfying
\beq\label{periodicitywx}
\Phi_w(x+2\pi R)=\Phi_w(x).
\eeq
It is of course tempting to interpret $\Phi_w(x)$ as a field in an effective space-time description. Clearly though, this is not really a local field in space-time in the usual sense, being at best an infinite set of fields labelled by $w$. 
At finite radius of compactification, we must keep the entire tower of such fields at hand. 

Interpreting it as a wave-functional, it is convenient to use the notation
\beq\label{Phiwx}
\Phi_w(x) \equiv 
\langle x,w|\Phi\rangle.
\eeq
What we will show here is that this notation is particularly effective, in that $\langle x,w|$ should be thought of as a {\it choice of basis} (the one given to us by the usual interpretation of the string zero modes as we have been describing here), and furthermore that by accounting for the non-commutativity of the zero mode sector, T-duality can be understood simply as a change of basis. The basis $| x,w\rangle$ corresponds to diagonalizing $\hat{x}$ and $\hat\tp$, which is a consistent choice, given that they commute (\ref{xpcomm}). 
Another basis which is commonly used to describe such states is the momentum basis which simultaneously diagonalizes
$\hat{p}$ and $\hat{\tp}$.
We introduce a ground state $|0,0\rangle$ annihilated by both and we define
\beq\label{mombas}
\langle n,w | := \langle 0,0 |e^{-in\hat{x}/R}e^{-iw\hat{\tx}/\tR}   
\eeq
There is an important subtlety inherent in this notation given the commutation relation (\ref{xtxcomm}) and as a result, operator ordering must be carefully managed. In particular we see that for the momentum basis we have to choose, as we have done in (\ref{mombas}), an order between the position and dual position.
Because of this operator ordering issue, we insist on a strict order for the labels on the basis states and always position the operators associated with $x$ and its momenta before those associated with $\tx$ and its momenta. As we will see,  T-duality reverses this order and 
this subtlety will lead directly to a well-known phase \cite{Frohlich:1993es,Hellerman:2006tx} in the effect of T-duality on states.

Returning to (\ref{Phiwx}), we have
\beq
\Phi_w(x) \equiv 
\langle x,0|e^{-iw\hat{\tx}/\tR}|\Phi\rangle.
\eeq
The states $\{ |w\rangle\}$ form a complete orthonormal basis,
 as do $\{ |n\rangle\}$. Assuming the normalization
$\langle x|n\rangle = e^{in x/R}$, we insert the identity
\beqn
\Phi_w(x)&=&
\sum_n \langle x,0|e^{in\hat{x}/R}|0,0\rangle\langle 0,0|e^{-in\hat{x}/R}e^{-iw\hat{\tx}/\tR}|\Phi\rangle 
\\&=&
\sum_n e^{inx/R}\langle n,w|\Phi\rangle 
\\&\equiv& \sum_n e^{inx/R} \Phi(n,w).\label{dualField}
\eeqn
In the last line, we have defined $\Phi(n,w)\equiv \langle n,w|\Phi\rangle $.
Clearly $\Phi(n,w)$ can be regarded as a state of fixed momentum and winding and (\ref{dualField}) can be regarded as a Fourier series.

What is perhaps not obvious is that we can also describe the same states in a dual basis, using basis states  $|n,\tx\rangle$, diagonalizing $\hat{p}$ and $ \hat{\tx}$. We interpret   $\Phi_n(\tx) = \langle n,\tx|\Phi\rangle$ as a collection of fields living in the dual space, and their periodicity
\beq\label{periodicityntx}
\Phi_n(\tx+2\pi \tilde{R})=\Phi_n(\tx),
\eeq
implies that $n$ can be interpreted as winding in the dual space.
 In fact, $\Phi_n(\tx)$ should be thought of as the image of $\Phi_w(x)$ under T-duality. Indeed, going from $\Phi_w(x)$ to $\Phi_n(\tx)$ corresponds to taking the data $(w,R;n,\tR)$ to $(n,\tR;w,R)$. We have
\beqn
\Phi_n(\tx) 
&=&\langle 0,\tx|e^{-in\hat{x}/R}|\Phi\rangle
\cr
&=&\sum_w\langle 0,\tx|e^{iw\hat{\tx}/\tR}|0,0\rangle\langle 0,0|e^{-iw\hat{\tx}/\tR}e^{-in\hat{x}/R}|\Phi\rangle
\cr
&=&\sum_we^{iw{\tx}/\tR} e^{i\pi nw} \langle 0,0|e^{-in\hat{x}/R} e^{-iw\hat{\tx}/\tR}|\Phi\rangle 
\cr
&=&
\sum_w e^{iw{\tx}/\tR}e^{i\pi nw}\Phi(n,w). \label{Field}
\eeqn
Thus $\Phi_n(\tx)$ is obtained from $\Phi(n,w)$ by a modified Fourier series containing an extra phase $e^{i\pi nw}$. The phase arises from the reorganization of  the order of the phase operators  in order to re-express the functional in  the momentum basis. Indeed, the non-trivial commutator (\ref{xtxcomm}) implies that
\be
 e^{-iw\hat{\tx}/\tR}e^{-in\hat{x}/R}  = e^{i \pi n w \frac{2\lambda^2}{R\tilde{R}}} e^{-in\hat{x}/R} e^{-iw\hat{\tx}/\tR} = e^{i \pi n w } e^{-in\hat{x}/R} e^{-iw\hat{\tx}/\tR}.
\ee 
Previously, such a phase has been uncovered at the level of states through a careful analysis of operatorial co-cycles \cite{Frohlich:1993es, Hellerman:2006tx}. Here, we see that it can be obtained in a straightforward way by instead taking into account the non-commutativity of $x$ and $\tx$. 
The important point is that the tower of fields $\Phi_w(x)$ contains the same information as the dual tower
$\Phi_n(\tx)$. We can express this equivalence directly by composing the relations (\ref{Field},\ref{dualField}).
Concretely, $\Phi_n(\tx)$ is related to $\Phi_w(x)$ as follows\beqn
\Phi_n(\tx)=\sum_{w\in \mathbb{Z}}e^{iw\tx/\tR}\int_0^{2\pi R} \!\!\frac{dx}{2\pi R} \, e^{-in(x-\pi wR)/R}\Phi_w(x) .\label{dblFour}
\eeqn
which we refer to as a double Fourier transform (or more properly, a Zak transform \cite{Freidel:2016pls}).

This formula should be regarded as the general statement of T-duality. Indeed, at large $R$, it is most natural to describe states in terms of the coordinate $x$ and winding $w$. On the other hand, at small $R$ (large $\tR$), it is most convenient to describe the states in terms of $\tx$ and $n$, interpreted now as dual winding. One can check that the transform does correctly localize onto $w=0$ for large $R$ and $n=0$ for large $\tR$. 
We also see that the tower $\Phi_n(\tx)$ contains precisely the same information as the tower $\Phi_w(x)$, and thus the coordinate $\tx$ plays a complementary role to that of $x$: one should simply choose one or the other, depending on the physics that one wishes to describe, the difference being simply a change of basis.  

We have seen that the non-trivial commutation relation results in a twist in the double Fourier transform -- there is an apparent half translation in $x$ (or equivalently in $\tx$). In fact, the double Fourier transform (\ref{dblFour}) is equivalent to the idea \cite{Freidel:2013zga, Freidel:2014qna} that T-duality itself can be regarded as a Fourier transform in the Polyakov path integral. Eq. (\ref{dblFour}) then is simply that relation obeyed by the Polyakov path integral, reduced to the zero mode sector. The extra phase comes about in that reduction from the aforementioned $e^{i\int\omega}$ factor in the Polyakov path integral. {\it We will thus express this by saying that T-duality is accompanied by a $\pi$-flux of $\omega$. }

Finally, let us define, by Fourier series, the generalized field that depends on two commuting labels $(x,\tx)$
\beq\label{Fieldd}
\Phi(x,\tx)\equiv \sum_w   \Phi_w(x)  e^{iw\tx/\tR} .
\eeq
Naively, it appears that this could be interpreted as a function on the double space. However, care must be taken in interpreting this object because of the underlying non-commutativity. Indeed, one finds that the above formulas imply that we can also write
\beq\label{Tfield}
\Phi(x,\tx)
&=&\sum_n e^{inx/R} \Phi_n(\tx-\pi n\tR) ,
\eeq
the half-shift coming from the extra phase in (\ref{dblFour}). 
Here we see that if we insist on keeping an interpretation where $x$ and $\tx$ are just commuting labels then T-duality appears as a non-local map. It is not just an exchange of $(n,R)\leftrightarrow(w,\tilde{R})$ and $x\leftrightarrow \tx$, but also involves arbitrarily large shifts in the dual variable. This basic fact shows that the double field theory interpretation of T-duality in terms of generalized fields as simply an exchange of $x$ with $\tx$ is not tenable.
We will come back to the proper interpretation of this wave functional later in the paper and resolve this puzzle.

\section{The Closed String Non-commutative Product}

\begin{figure}[!h]\centering
\subcaptionbox{{\small String product}}[0.3\textwidth]{
	\includegraphics[width=3.5cm]{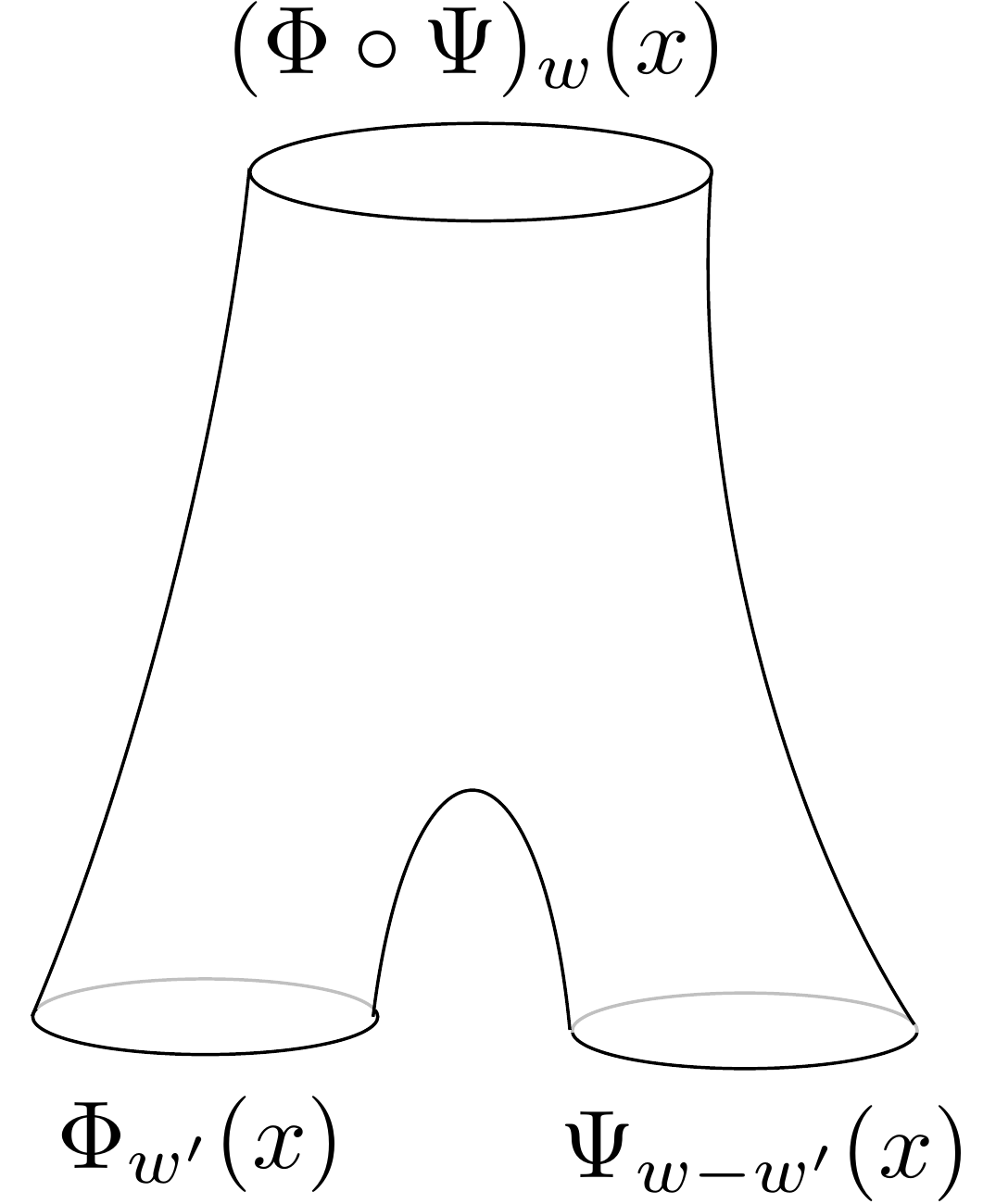}\label{fig:pants}
	}
\subcaptionbox{{\small Nakamura cutting}}[0.3\textwidth]{
	\includegraphics[width=3.5cm]{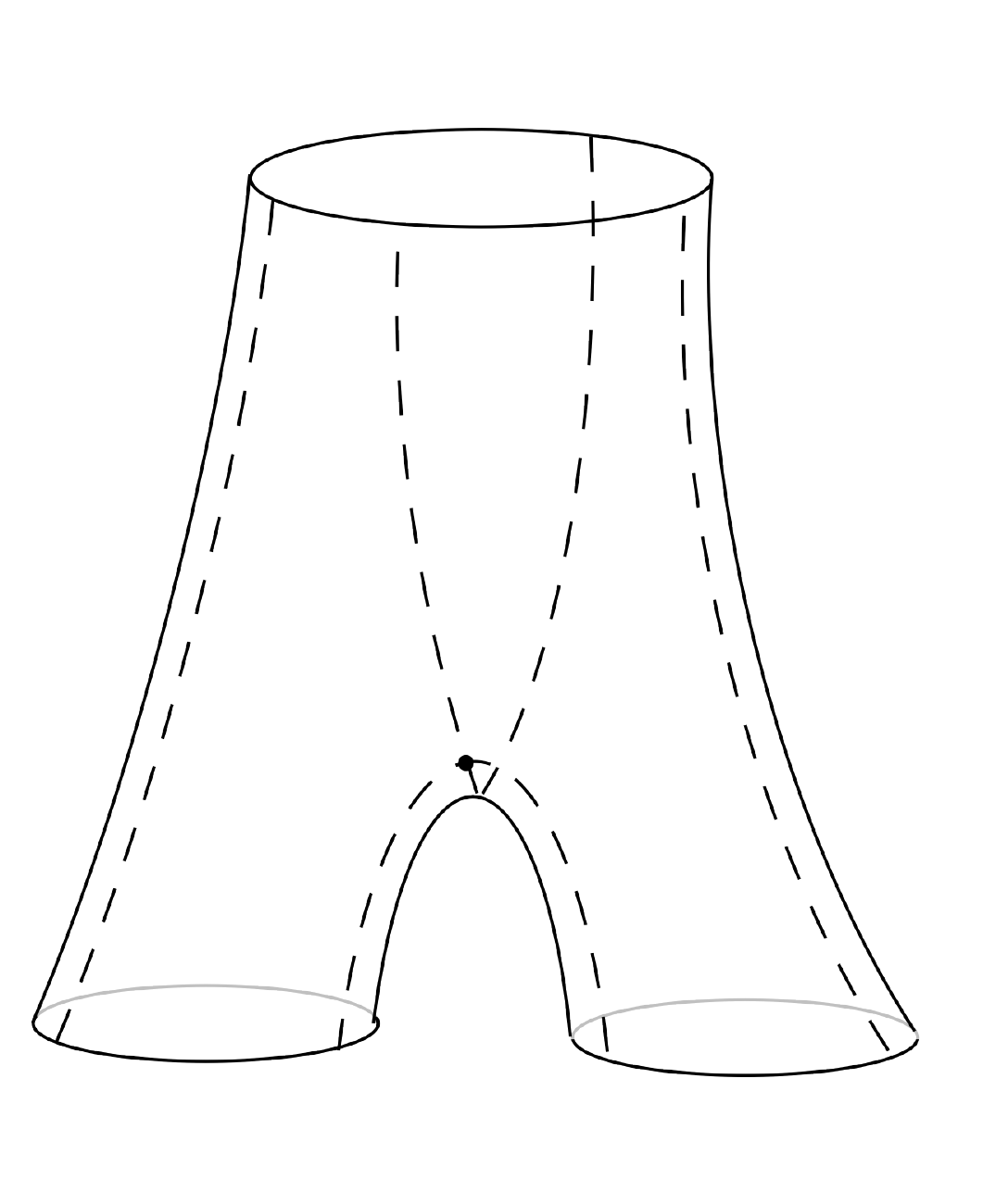}\label{fig:pants-Nakamura}}	
\subcaptionbox{{\small unfolded along cuts}}[0.3\textwidth]{
	\includegraphics[width=3.5cm]{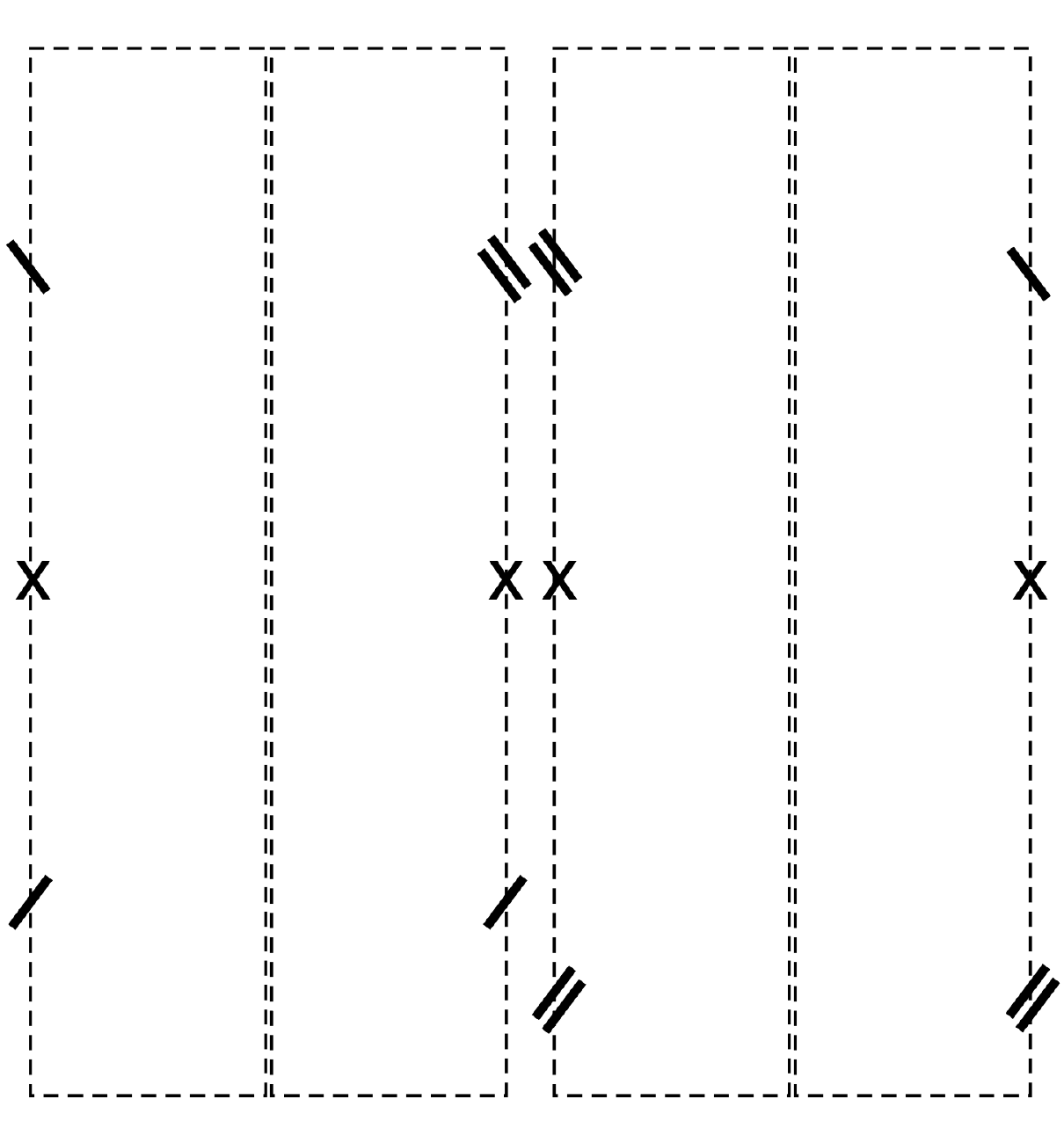}\label{fig:pants-Nakamura}}	
\caption{{\small The worldsheet pants diagram for states written in $w,x$ basis. We map to the covering space of the target space, in which $\Phi_w(x)$ represents a string extended from $x$ to $x+2\pi wR$. To arrange this interaction, we split this string at its midpoint into pieces that we denote $\Phi^{(+)}$ and $\Phi^{(-)}$, and affect the rejoining at the midpoint. }}
\label{fig:pants1}
\end{figure}

The pants diagram of perturbative closed string theory can be interpreted as defining a product of closed strings corresponding to the splitting-joining interaction, as drawn in Fig. \ref{fig:pants1}.

We wish to investigate what the non-commutativity implies for this product. We will claim that the non-commutativity offers a precise (and unique) way to interpret a closed string as decomposable into a pair of `open strings', fully consistent with T-duality.

\begin{figure}[!h]\centering
	\includegraphics[width=9cm]{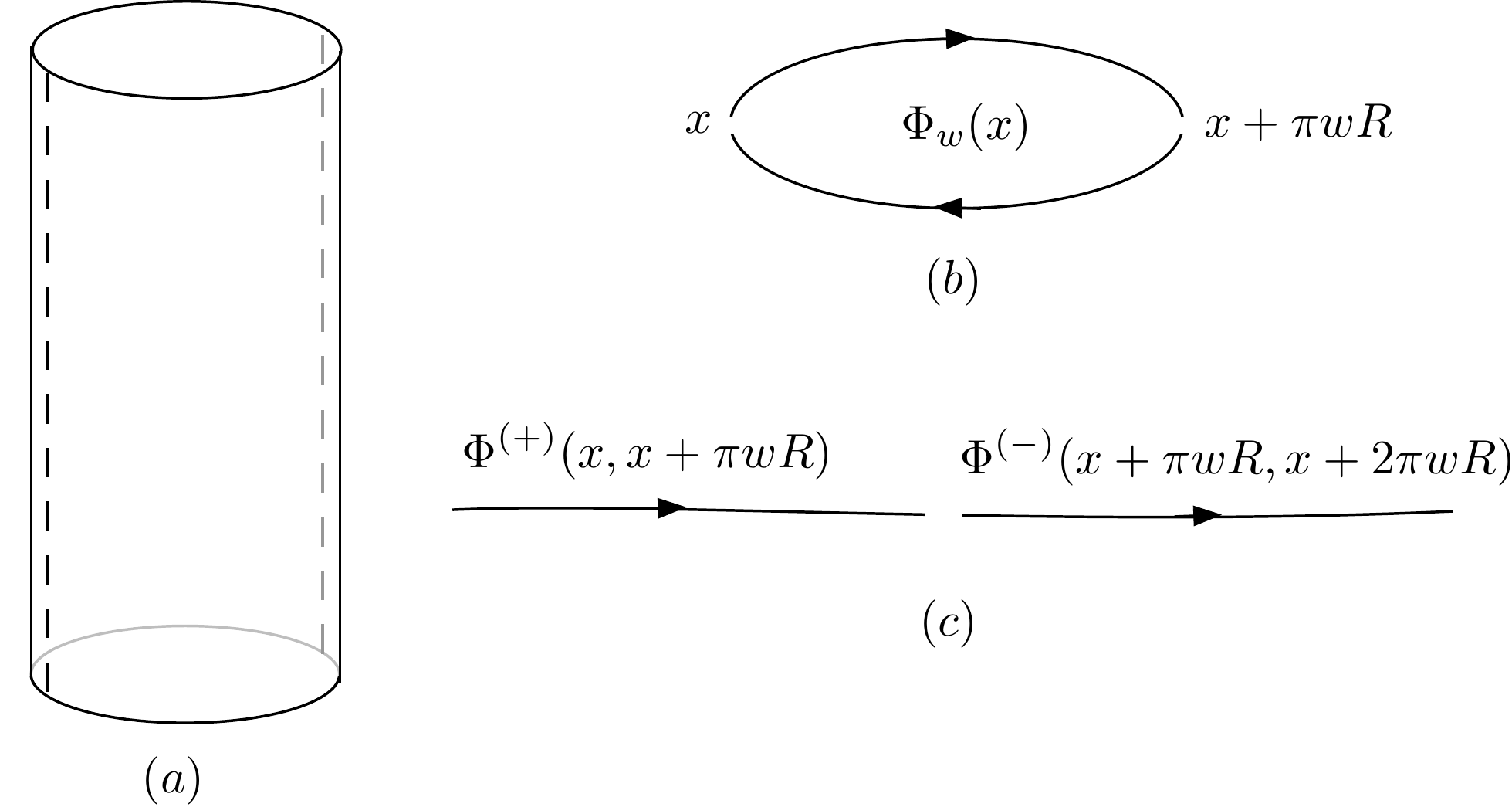}
	\caption{Closed string splitting: (a) the worldsheet is sliced open at, say, $\sigma=0,\pi$. (b) The corresponding embedding is described by $\Phi_w(x)$, now separated into two pieces. (c) The pieces of the closed string unfolded onto the covering space.}\label{fig:splitstring.pdf}
\end{figure}

Indeed, to describe the splitting-joining interaction, it is convenient to first split the closed strings into two half strings, which are glued in the middle. We will find that where this is done in the embedded space is dictated by the strength of the commutator (\ref{xtxcomm}): it must be done precisely  at the midpoint in the target space. The splitting is shown graphically in Fig. \ref{fig:splitstring.pdf} and we will denote it as an ordered product\footnote{Note that here we are using notation that might imply, in given worldsheet coordinates, $x=x(\sigma=0)$, $x(\sigma=2\pi)=x+2\pi wR$. This is for convenience only (it is precise only if we neglect the oscillators). What we are describing as the mid-point is in fact the center of mass position, $\frac{1}{2\pi}\int_0^{2\pi} d\sigma  X(\tau,\sigma) = x(\tau)+\pi w R$.  }
\beq
\Phi_w(x) = \Phi^{(+)}(x,x+\pi wR)\Phi^{(-)}(x+\pi wR,x+2\pi wR) .
\eeq

\noindent In other words, the closed string field is regarded as the product of two half-string fields $\Phi^{(\pm)}$ of target length $\pi wR$, with $\Phi^{(-)}$ translated by a distance $\pi wR$. 
An important property that we demand for the half-string fields is that they respect the lattice periodicity condition. If one translates the initial and final point of the string by a lattice distance nothing changes
\be
\Phi^{(\pm)}(x+2\pi R, y+2\pi R)= \Phi^{(\pm)}(x,y). 
\ee
The splitting-joining interaction is obtained by performing this splitting on each closed string, by multiplying each half string  and then  rejoining at the midpoint, as we show in Fig. \ref{fig:splitjoinstring.pdf}.\footnote{
The Nakamura cutting \cite{nakamura2000calculation} is the most natural to construct the splitting-joining interaction. We note in passing that in Lorentzian signature, one can choose a metric that is everywhere Lorentzian except at isolated points where worldsheet curvature singularities arise --- intuitively, the curvature singularity is where the dilaton couples. In the case of the pants diagram, one can choose this to occur at a single point, denoted by an {\sffamily x} in Fig. \ref{fig:pants1}(c). As we are about to show, the pants vertex is also associated with a certain $\pi$-flux. Recently in another context, the cutting of the pants into a pair of `hexagons' has been considered\cite{Basso:2015zoa}. It is not clear to us how to establish that the gluing of hexagons should be accompanied by this $\pi$-flux, or what relevance it might have in the context of the cited reference.
}
\myfig{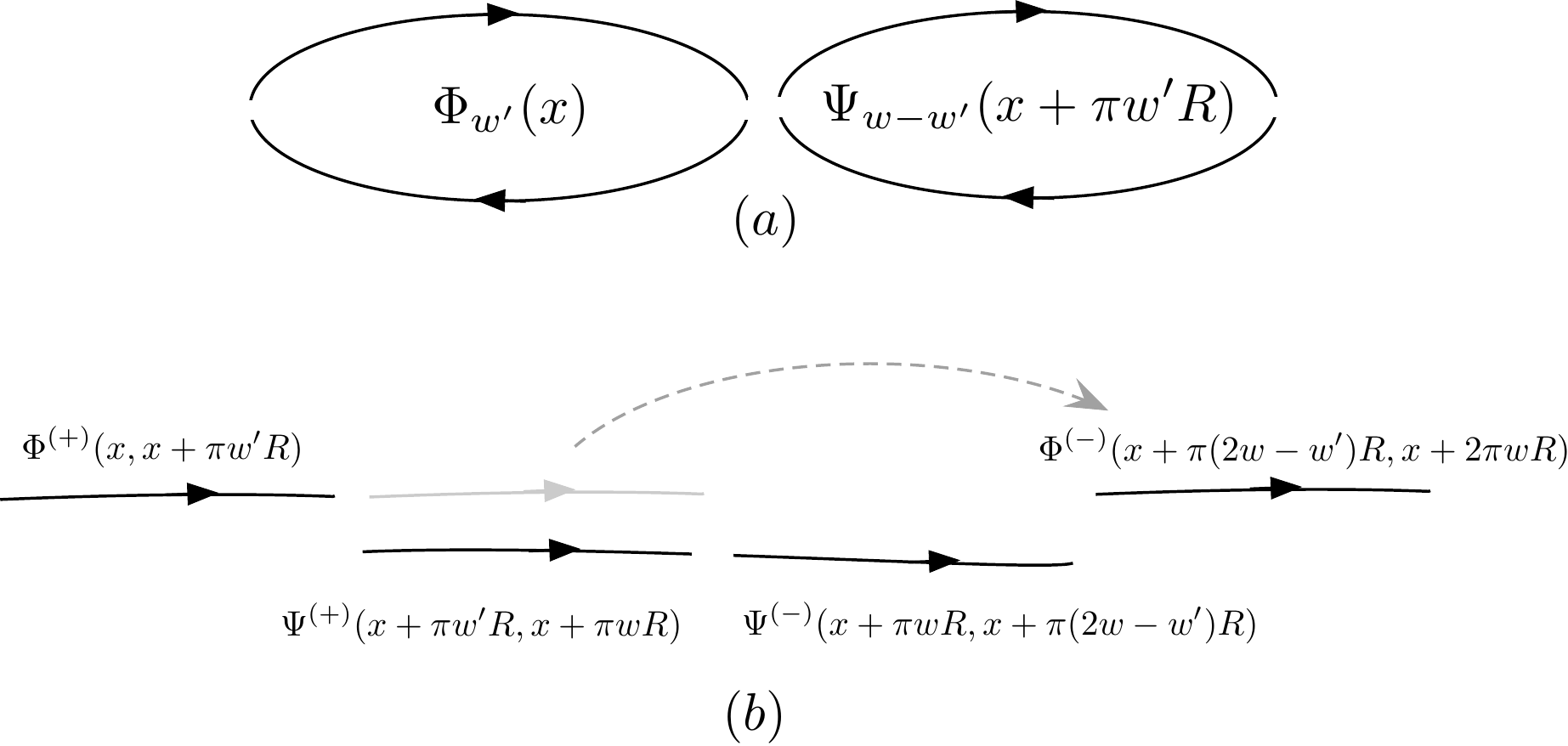}{10}{The splitting-joining interaction is effected by splitting each string at its midpoint and reattaching. In the covering space, this process involves translating half of one of the strings ($\Phi^{(-)}$) by a lattice vector $2\pi R(w-w')$.}

This procedure can be formalized as follows. First one introduces the half string product which is just the open string concatenation
\be 
(\Phi\circ\Psi)^{(\pm)}_{w'}(x,x+\pi wR)\equiv \Phi^{(\pm)}(x,x+\pi w'R)\Psi^{(\pm)}(x+\pi w'R,x+\pi w R).
\ee
This simply corresponds to rejoining the string segments at the ends which are at the same point in the target, similar to the product introduced by Douglas and Hull \cite{Douglas:1997fm}. 
Then we can construct the closed string product by splitting and rejoining as in Fig. \ref{fig:splitjoinstring.pdf}.
This procedure, shown in Fig. \ref{fig:splitjoinstring.pdf}(b) for fixed $w'$ and $w$, corresponds then to the product
\beq
(\Phi \circ \Psi)^{(+)}_{w'}(x;x+\pi wR) 
(\Psi \circ \Phi)^{(-)}_{w'}(x+\pi wR; x+ 2\pi wR) 
\eeq
which is given by
\beqn
&& \Phi^{(+)}(x;x+\pi w'R)\Phi^{(-)}(x+\pi w'R;x+2\pi w'R)\\
&&\times\Psi^{(+)}(x+\pi w'R;x+\pi w R)\Psi^{(-)}(x+\pi wR;x+\pi w'R+2\pi (w-w')R)\nonumber\\
&=& \Phi_{w'}(x)\Psi_{w-w'}(x+\pi w'R) 
\eeqn
 We then define the {\it closed string product}, with total winding $w$, as the sum over all the intermediate windings 
\beqn
(\Phi\circ\Psi)_{w}(x)&=& \sum_{w'}(\Phi\circ\Psi)_{w'}^{(+)}(x;x+\pi wR)(\Phi\circ\Psi)_{w'}^{(-)}(x+\pi wR;x+2\pi wR)\label{sj3}
\\
&=&\sum_{w'\in\mathbb{Z}}\Phi_{w'}(x)\Psi_{w-w'}(x+\pi w'R) .
\eeqn
It is not at all obvious that the splitting-joining interaction should take place at precisely the midpoint in the target space. What we find is that this choice corresponds precisely to the presence of the non-commutative phase! Furthermore, it is only for this specific non-commutative phase that the string interaction is consistent with T-duality. 

To understand these claims, it is instructive to consider the transform to the doubled space
\beqn
(\Phi\circ\Psi)(x,\tx)&=&\sum_w e^{iw\tx/\tR} (\Phi\circ\Psi)_{w}(x)\cr
&=& \sum_{w,w'} e^{iw\tx/\tR}\Phi_{w'}(x)\Psi_{w-w'}(x+\pi w'R)\nonumber\\
&=&\sum_{w'} e^{iw'\tx/\tR}\Phi_{w'}(x)\Psi(x+\pi w'R,\tx) .
\eeqn
We caution the reader again (see the discussion at the end of Section \ref{sec:Tdual}) that although the notation implies a function on the space coordinatized by $x,\tx$, there is a sublety here. Indeed, in order to understand the nature of this string product, let us specialize to the case in which $\Phi(x,\tx)=\varphi(x)$ and $\Psi(x,\tx)=\psi(\tx)$, and we will compare  $(\Phi\circ\Psi)$ to $(\Psi\circ\Phi)$. Further, take plane waves $\varphi(x)=e_n(x)=e^{inx/R}$ (a pure momentum mode) and $\psi(\tx)=\tilde{e}_w(\tx)=e^{iw\tx/\tR}$ (a pure winding mode). Corresponding to these plane waves are fields $\Phi_{\tilde{w}}(x)=\delta_{\tilde{w},0}e^{inx/R}$ and  $\Psi_{\tilde{w}}(x)=\delta_{w,\tilde{w}}$.
Then we have
\beqn
(e_n\circ\tilde{e}_w)(x,\tx)
= \sum_{\tilde{w},w'} e^{i\tilde{w}\tx/\tR}\delta_{w',0}e^{inx/R}\delta_{\tilde{w},w-w'}
=e^{iw\tx/\tR}e^{inx/R}=e_n(x)\tilde{e}_w(\tx) ,
\eeqn
and
\beqn
(\tilde{e}_w\circ e_n)(x,\tx)
=\sum_{\tilde{w},w'} e^{i\tilde{w}\tx/\tR}\delta_{w,w'}\delta_{\tilde{w},w'}e^{in(x+\pi w'R)/R}
=e^{i\pi nw}e^{iw\tx/\tR}e^{inx/R}
=e^{i\pi nw}e_n(x)\tilde{e}_w(\tx).
\eeqn
This suggests that the string product can be interpreted as a realization of the non-commutative product  between operators $\hat{x}$ and $\hat{\tx}$. The non-commutativity 
\beq
 (\tilde{e}_w \circ e_n )= e^{i\pi n w} (e_n\circ \tilde{e}_w),
\eeq
  captured by the string product  is an instance of the Heisenberg group; the strings can be thought of in terms of the corresponding operatorial representation ($x\to \hat x, \tx\to \hat{\tx}$) and the string interaction is given by the Heisenberg product implied by the commutation relation 
\beq\label{xtxcomm1}
[\hat{x},\hat{\tx}]=i\pi R\tR=2\pi i\lambda^2.
\eeq
In fact, the actual numerical value of the commutator (\ref{xtxcomm1}) is correlated with the point of attachment of the half-strings in the splitting-joining interaction. We believe that this result gives a precise sense in which closed strings can be thought to decompose into a pair of open strings. Associated closely with this result though is the non-commutativity of the zero modes.

We can also express the string product in momentum-winding space. A short calculation yields
\beqn
(\Phi\circ\Psi)(n,w)
&=&\sum_{w',n'}\Phi({n',w'})  e^{i\pi w' (n-n')}\Psi(n-n',{w-w'}). \eeqn
Of course, we find that this product expresses the conservation of both momentum and winding under multiplication since the total momenta $n$ of $(\Phi\circ\Psi)$ is given by the sum of individual momenta associated with $\Phi$ and $\Psi$ and similarly for the winding. In addition,  we find that  the vertex comes with a $\pi$-flux phase which is the Fourier transform of the non-commutativity of the product. It also expresses, as we will see, the presence of a symplectic flux $\int \omega$ in the string interaction.

In the construction of the string product we have chosen the re-attachment point to be the mid-point in the target of the winding string. This might look like a very symmetric but somewhat arbitrary choice. It turns out that this choice  is the 
only choice of string product consistent with T-duality.
 Of course, it is well known that T-duality is not only a property of the spectrum of the theory but extends also to interactions. In our context, we take this to mean that the string interaction transforms consistently under T-duality via the double Fourier transform. Indeed, one can show that the double Fourier transform of the product is
\beq
\sum_{w}e^{iw\tx/\tR}\int \!\!\frac{dx}{2\pi R} \, e^{-in(x-\pi wR)/R}\, (\Phi\circ \Psi)_w(x)
= \sum_{n'} \Phi_{n'}( \tx ) \Psi_{n-n'}(\tx+ \pi n' \tR)
= (\Phi\circ \Psi)_n(\tx) .
\eeq
That is the transform of the product equals the product of the transforms.\footnote{As we have stated above, one can prove that this is if and only if we attach at the mid-point.}

%
%

\section{Non-commutativity and fluxes}

As promised earlier, we now return to discuss the  proper interpretation of the generalized field 
\beq\label{Gfield}
\Phi(x,\tx)\equiv \sum_w \Phi_w(x) e^{iw\tx/\tR}.
\eeq 
As we have seen in the previous section the string product is essentially a representation of the Heisenberg group. This suggest that one should  consider the ``quantization'' map 
\beq\label{Gfieldw}
\Phi(x,\tx) \to \hat{\Phi} =\sum_w \Phi_w(\hat{x}) e^{iw\hat{\tx}/\tR},
\eeq from generalized fields to  non-commutative fields.\footnote{Here, we have chosen a specific operator ordering. Given this ordering, the mapping is well-defined and consistent with the string product.}
This map possesses two key properties: first, under this map the T-duality transformation (\ref{Tfield}) becomes ``localized'' and is expressed as the exchange of $\hat{x}$ with $\hat{\tilde{x}}$. Indeed, using the periodicity (\ref{periodicityntx}),  the T-dual expression is given by
\beq\label{Tfield}
\hat\Phi=\sum_n e^{in\hat{x}/R} \Phi_n(\hat{\tx}-\pi n\tR)
=
\sum_n   \Phi_n(\hat\tx) e^{in\hat{x}/R},
\eeq 
which has a similar form to (\ref{Gfieldw}).
We see that the non-commutativity of $\hat{x}$ with $\hat\tx$ allows one to reabsorb all the shifts in terms of a simple reordering that exchanges  $\hat{x}$ with $\hat\tx$ and is the expression of T-duality.  The ``quantized'' field is simply expanded in terms of modes as 
\beq
\hat\Phi \equiv \sum_{w,n}  e^{in\hat{x}/R}  \Phi(n,w) e^{iw\hat{\tx}/\tR}. 
\eeq
The second property  is that the quantization map  also ``localizes'' the string product:
\beq
\widehat{\Phi \circ \Psi} = \hat\Phi \hat\Psi. 
\eeq

It is useful at this point to generalize the construction to higher dimensional tori. This can be done in a straightforward manner by introducing the modes
$\K^A=(\tk^a, k_a )$, generalizing  $ ( w/\tilde{R},n/R)$. We also introduce a ``para-hermitian'' structure $(\eta,\omega)$ where 
\beq\label{flatetaom}
\eta(\K,\K') =k \cdot \tilde{k}' + \tilde{k}\cdot k',\qquad
\omega(\K,\K')=k \cdot \tilde{k}' - \tilde{k}\cdot k'.
\eeq
The integrality condition for the lattice $\Lambda$ of admissible modes $\K,\K'\in\Lambda$ reads in this notation as\footnote{In the one dimensional case where $\K=(w/\tilde{R},n/R)$ this follows directly from 
$(\eta+ \omega)(\lambda \K,\lambda \K') = n w'$ and similarly $(\eta- \omega)(\lambda \K,\lambda \K')= wn' $, given that $n,n',w,w'\in\mathbb{Z}$.}
\beq\label{int}
(\eta \pm \omega)(\lambda \K,\lambda \K')\in \mathbb{Z}.
\eeq
Recall that when we introduced $\Phi(n,w)$ above, we were led to insist on an ordering for the labels. In the present case, we now write $\Phi(\K)=\langle \K|\Phi\rangle$ with the  ordering chosen as
\be
\langle \K| = \langle 0| \hat{U}_{-\K},\qquad \hat{U}_{\K} \equiv e^{i k\cdot \hat{x}} e^{i\tilde{k} \cdot \hat{\tx}}.
\ee
This ordering can be seen to be related to the choice of an O$(d,d)$ frame, where we place the operator associated with ${x}$ on the left and the operator associated with the dual space $\tilde{x}$ on the right. 
The key point is that this choice of frame is entirely encoded into the choice of symplectic potential $\omega$ and the previous wave operator can be covariantly written in terms of $\K=(\tilde{k},k) $ and $\X=(x,\tilde{x})$ as 
\be 
\hat{U}_{\K} = e^{\frac{i}{2} (\eta + \omega)(\K,\hat\X)}e^{\frac{i}{2} (\eta - \omega)(\K,\hat\X)}.
\ee
Given this notation we can write the string product covariantly as 
\beq \label{proddual}
(\Phi\circ\Psi)(\K)
&=&\sum_{\K'+\K'' =\K }  \Phi(\K') e^{ i\pi(\eta- \omega)(\lambda\K',\lambda\K'')}\Psi(\K'').
\eeq
The non-commutativity of the string product is encoded in terms of a $\pi$-flux due to $\omega$. As it turns out the phase factor is exactly the same as the cocycle factor
$\epsilon(\K,\K')= e^{ i\pi (\eta- \omega)(\lambda\K,\lambda\K')}$ that appears in the 
definition of the vertex operator product \cite{Polchinski:1998rq,Freidel:2017wst}.\footnote{
This is of course a straightforward consequence of the relationship between the sphere amplitude with three insertions and the pants diagram with fixed states on each leg. The result is a strong and pleasing indication of consistency.}
We can also introduce the  generalized fields $\Phi(\X)$ and the correspond quantized operator $\hat\Phi$ as follows
\be\label{Fields}
 \Phi(\X) =\sum_{\K \in \Lambda} e^{i \eta(\K,\X)} \Phi(\K) ,\qquad
 \hat\Phi = \sum_{\K \in \Lambda} e^{\frac{i}{2} (\eta + \omega)(\K,\hat\X)}\Phi(\K) e^{\frac{i}{2} (\eta - \omega)(\K,\hat\X)}.
\ee
We see that $\omega$ enters the quantum field definition in the choice of operator ordering. 
The  product of quantum operators defines a star-product on the generalized  fields  defined by $\widehat{\Phi \circ \Psi} \equiv \hat\Phi \hat\Psi$ and which is given explicitly by
\be \label{prod}
(\Phi \circ_\omega \Psi)(\X) 
=m \left(  e^{2\pi i\lambda^2 (\tilde{\pa}^a\otimes  \pa_a) }  \Phi(\X) \otimes\Psi(\X) \right).
\ee
where $m$ denotes the pointwise multiplication $ m(\Phi(\X) \otimes\Psi(\X)) = \Phi(\X)\Psi(\X)$.
It is interesting to note that the section condition $\pa_A\Phi\pa^A\Psi=0$ imposed in double field theory  implies that the string product reduces to the commutative pointwise product
$(\Phi \circ \Psi)(\X)  = \Phi(\X)  \Psi(\X) $.

In summary, the  fields that enter the effective description of the  string compactified on a  d-dimensional torus are functions on the 2d-dimensional
torus $\mathbb{T}_\Lambda= \mathbb{C}^{2d}/\Lambda$. In other words, the generalized fields are periodic, with period $\Lambda$. 
This space of fields is equipped with a non-commutative product $\circ_\omega$ which depends on the symplectic structure and defines a non-commutative algebra
\be
\mathbb{A}_{\Lambda,\omega}= (  C^{\infty}(\mathbb{T}_\Lambda) , \circ_\omega),
\ee
which is a multidimensional non-commutative torus \cite{Rieffel}. A very important point about this algebra is that although it is non-commutative, it possesses a very large center, that is, it is {\it almost} commutative. 
This is due to the fact that the non-commutativity is due to $\pi$-flux. The center of $\mathbb{A}_{\Lambda,\omega}$ is simply associated with the double lattice $2\Lambda$.
It is indeed clear from (\ref{proddual}) and the condition (\ref{int}) that fields whose mode function $\Phi(\K)$ vanishes, unless $\K \in 2 \Lambda$, form a subset of fields that commute with any other periodic field. The center fields are in $C^{\infty}(\mathbb{T}_{\Lambda /2})
$ and satisfy the stronger periodicity condition 
\be
\Phi(\X+\K/2)= \Phi(\X), \qquad \K \in \Lambda. 
\ee 
The center algebra is an example of a {\it modular algebra}, {\it i.e.}, a commutative algebra embedded  in a non-commutative algebra which has no classical analog \cite{Freidel:2016pls, Aharonov:1969qx}. 

\subsection{$O(d,d)$ and Non-trivial Constant Backgrounds}

So far we have assumed that the background is trivial, with the fields $(\eta,\omega)$  {\it constant} and given by (\ref{flatetaom}). 
As shown in \cite{Freidel:2017wst}, we can turn on non-trivial backgrounds encoded into $\omega$ by changing the O$(d,d)$ frame $\X \to O\X$. This change of frame preserves $\eta$ but  transforms $\omega$. Any constant $\omega$ can be obtained this way. Since $\omega$ has an interpretation as the symplectic form on the space of $\X$'s, modifying $\omega$ affects the commutation relations\footnote{The algebraic structure that we are working with here has an analogy in electromagnetism in the presence of monopoles. In that analogy, the string length becomes the magnetic length, and the form $\omega$ becomes the magnetic field.  Another analogy occurs in quantum Hall liquids, the algebra being the magnetic algebra of the lowest Landau level.
}
\be
[\hat{\X}^A,\hat{\X}^B]= 2\pi i\lambda^2\Pi^{AB},\qquad \Pi^{AB} \omega_{BC}= \delta^A{}_C, 
\ee
where we have introduced the Poisson tensor $\Pi=\omega^{-1}$. 

 For instance, under a constant $B$-field transformation 
 $\X=(x^a,\tx_a) \mapsto (x^a, \tx_a + B_{ab} x^b)$, 
 the trivial symplectic form (\ref{flatetaom}) is mapped onto
$
\omega(\K,\K')=  k_a  \tk'^a - k'_a \tk^a - 2 B_{ab} \tk^a \tk'^b,
$
and the commutators read
 \beq\label{Bcommrel}
[\hat{x}^a,\hat{x}^b]=0,\qquad 
[\hat{x}^a,\hat{\tx}_b]=2\pi i\lambda^2 \delta^a{}_b,\qquad 
[\hat{\tx}_a,\hat{\tx}_b]=-4\pi i\lambda^2 B_{ab}.
\eeq
We see that the  effect of the $B$-field is to render the dual coordinates non-commutative. More generally, we can parameterize an arbitrary $O(d,d)$ transformation as $g=e^{\hat{B}}\hat{A}e^{\hat\beta}$, where $\hat {A}\in GL(d)$ and $e^{\hat{B}}=\tiny\begin{pmatrix}{\bf 1}&0\cr B & {\bf 1}\end{pmatrix}$ and $e^{\hat{\beta}}=\tiny\begin{pmatrix}{\bf 1}&\beta\cr {\bf 0} & {\bf 1}\end{pmatrix}$ are nilpotent. $e^{\hat B}$ is the $B$-field transformation discussed above, and is associated with the usual $B$-field deformation in string theory. We note that the transformation of $(x^a,\tx_a)$ given above does not modify $x^a$, and thus fields that depend only on $x^a$ are unmodified. The $\beta$-transformation on the other hand corresponds to the map $(x^a,\tx_a)\mapsto (x^a+\beta^{ab}\tx_b,\tx_a)$. Equivalently, it has the effect of mapping the symplectic structure to
$
\omega(\K,\K')=  k_a  \tk'^a - k'_a \tk^a + 2 \beta^{ab} k_a k'_b,
$
and yields commutation relations
 \beq\label{betacommrel}
[\hat{x}^a,\hat{x}^b]=4\pi i\lambda^2 \beta^{ab},\qquad 
[\hat{x}^a,\hat{\tx}_b]=2\pi i\lambda^2 \delta^a{}_b,\qquad 
[\hat{\tx}_a,\hat{\tx}_b]=0.
\eeq
Dramatically, the coordinates that are usually thought of as the space-time coordinates have become themselves non-commutative. Since this is the result of an $O(d,d)$ transformation, we know that it can be thought of in similar terms as the $B$-field; these are related by T-duality. We are familiar with the $B$-field background because we have, in the non-compact case, a fixed notion of locality in the target space theory. However, in the non-geometric $\beta$-field background, we do not have such a notion of locality but access it through T-duality.

\subsection{Associativity and Flux}

We believe that the $\pi$-flux that we have displayed above is fundamental, and will persist to non-constant backgrounds. In that context then, the non-closure of $\omega$, that is $H=d\omega\neq 0$ will lead to a non-associative zero mode algebra 
\cite{Jackiw:1984rd, Jackiw:1985hq, Grossman:1984fs}. Indeed  one assumes that even in the presence of a non-trivial $B$-field that depends only on $x$ the commutation relations given above are preserved and we can easily check that the Jacobi identity is anomalous and given by 
\be
[\hat\tx_a,[\hat\tx_b,\hat\tx_c]]+ cycl.= H_{abc}(x)  ,
\ee
where $H_{abc}=\pa_a B_{bc} + cycl.$ is the H-flux. The relationship between the presence of flux in string theory  and non-associative geometry  has been discussed previously, in
\cite{Lust:2010iy, Blumenhagen:2010hj, Blumenhagen:2011ph, Blumenhagen:2011yv, Lust:2012fp, Andriot:2012vb, Gunaydin:2013nqa, Mylonas:2014aga}. Here we are seeing non-associativity
directly from the deformation of the symplectic structure of the zero modes.
As it has been recently argued in \cite{Freidel:2017yuv} the generalization of the geometry allows for an extension of the para-K\"ahler structure $({\cal P}, \eta,\omega)$ into a more general para-hermitian structure. In this extension we keep the condition 
that $K\equiv \eta^{-1}\omega$ is a split structure satisfying $K^2=1$ but we can relax the condition of closure and allow for a non-trivial flux ${\cal F} = \rd \omega$.
The existence of the  split structure $K$ admits a decomposition of  the tangent space of  $\cal P$ in terms of its eigenspaces which are Lagrangians. These Lagrangians play the role of the commutative subsets labeled by $x$ and $\tx$. The non-commutative product can then in principle be constructed from the knowledge of $\omega$ and a choice of para-hermitian connection \cite{Fedosov:1994zz, Fedosov}.

Here, ${\cal F}$ can be interpreted as a 3-form on ${\cal P}$, playing the role of a 3-cocycle and containing `non-geometric' fluxes, as it generally will have components of type $(3,0), (2,1), (1,2), (0,3)$ with respect to the coordinatization $(x,\tx)$.  These fluxes are respectively related to the $H$-flux, $F$-flux, and $R$-flux appearing in double field theory \cite{Shelton:2005cf, Hassler:2014sba,Aldazabal:2013sca}.
The relationship can be explicitly unraveled by introducing generalized frame fields and dual forms
\be
\hat{E}_I= \hat{E}_I{}^A \pa_A,\qquad E^I= \rd\X^A E_A{}^I,\qquad
\hat{E}_I{}^AE_A{}^J = \delta_I^J ,
\ee
where $\pa_A=(\tilde{\pa}^a,\pa_a)$ are derivatives on $\cal P$. The fully dressed 2-form $\omega$ can be expanded  as
\be
\omega = \omega_{IJ} E^I \wedge E^J,\qquad \eta = \eta_{IJ} E^I \otimes E^I, 
\ee
where $(\omega_{IJ},\eta_{IJ})$ are the constant two-form and metric  defined in (\ref{flatetaom}). All the $\X$ dependence is in the generalized frame $\hat{E}(\X)$. The frame is usually taken to be an element of O$(d,d)$, so that the metric $\eta$ is unchanged.\footnote{The possibility to relax this condition has been investigated in \cite{Freidel:2017yuv}.}
The Cartan structure equation defines a structure constant ${\cal C}_{IJ}{}^K$ given by 
\be
[\hat{E}_I,\hat{E}_J]={\cal C}_{IJ}{}^K \hat{E}_K,\qquad
\rd E^K +   {\cal C}_{IJ}{}^K E^I \wedge E^J=0,
\ee
and the symplectic structure can be expanded in terms of the flux
\be
\rd \omega= 2{\cal F}^\omega_{IJK}  E^I \wedge E^J\wedge E^K.
\ee
Here we have defined $ {\cal F}^\omega_{IJK} = 3{\cal C}_{[IJ}{}^D \omega_{K ]D}$. This should be compared with the 
usual flux defined in double field theory and given by 
$ {\cal F}^\eta_{IJK} = 3{\cal C}_{[IJ}{}^L \eta_{K] L}$.
They only differ by signs on the corresponding Lagrangian subspaces. 
This flux can then be expanded in components 
$(H_{abc}, F_{ab}{}^c,Q^{ab}{}_c,R^{abc})$ and contains the key information about the non-geometric backgrounds.
Here they appear as a parametrization of the non-trivial commutation relations and their lack of associativity.
The recent proposal of \cite{Freidel:2017yuv} is that the section condition $\eta^{AB}\pa_A\Phi\pa_B\Psi=0$ does not determine which sections one chooses and it should be supplemented by the Lagrangian condition $\omega^{AB}\pa_A\Phi\pa_B\Psi=0$. 
It is tantalizing to consider that imposing the section condition  with the help of $\omega$ implies a relationship between  ${\cal F}^\eta$ and ${\cal F}^\omega$.
A non-trivial question in this setting is to understand the nature of the generalization of the lattice $\Lambda$, presumably expressed in terms of parallel transport with respect to a para-hermitian connection preserving $\eta$ and $\omega$.

We conclude this section with a historical remark:
the fact that the string product 
is a representation of the Heisenberg algebra \cite{Heisenberg1925} is analogous to how the Heisenberg algebra was discovered in the original work of Born and Jordan \cite{BornJordan1926} that immediately followed that of Heisenberg \cite{Heisenberg1925}. 
Heisenberg showed that the Ritz law of composition of spectral frequencies $\nu_{ik}=\nu_{ij}+\nu_{jk}$ forces the composition of physical operators, such as the position of the electron, to satisfy a composition law similar to the string product.
Born and Jordan \cite{BornJordan1926} (and also \cite{Dirac1925, BornHeisenbergJordan1926}) realized that this implied an underlying  non-commutative structure encoded into the so-called Heisenberg algebra.
This nicely ties to our previous observations that the geometry of generic representations of quantum theory \cite{Freidel:2016pls} is realized in the metastring formulation of string theory \cite{Freidel:2015pka}. See also \cite{Bars:2014jca} for a direct relationship between the string field product and the Heisenberg algebra.
 It is interesting to note that the connection, discovered by Born and Jordan,  between the groupoid represented by the string product and non-commutative algebra is also at the heart of the field of non-commutative geometry of Connes \cite{Connes}.

\section{Conclusions}

In this paper we have presented several consequences of the non-commutativity of the zero mode sector in toroidal compactifications of  closed  string theory. It seems natural to suppose that there are further deep consequences, both for effective field theories as well as more formal aspects of string theory. We have in particular noted that there is a simple closed string product, equivalent to the splitting-joining interaction of the pants, that respects this non-commutativity as well as T-duality. Its structure is suggestive that
the non-commutativity of the zero mode sector will play an important role in the non-perturbative structure 
of the theory.

\bigskip

{\bf Acknowledgements:}
{\small RGL} and {\small DM} thank Perimeter Institute for
hospitality. {\small LF}, {\small RGL} and {\small DM} thank the Banff Center for
providing an inspiring environment for work and the Julian Schwinger Foundation for support. {\small RGL} is supported in part by
the U.S. Department of Energy contract DE-SC0015655 and
{\small DM}
by the U.S. Department of Energy
under contract DE-FG02-13ER41917. Research at Perimeter Institute for
Theoretical Physics is supported in part by the Government of Canada through NSERC and by the Province of Ontario through
MRI.

%

\providecommand{\href}[2]{#2}\begingroup\raggedright\endgroup

\end{document}